# NaCl-Assisted Growth of Ferroelectric SnSe Nanosheets with Spin Glass-like Behavior

Huiwen Xu[1, 2], Hanxiang Wu[1, 2], Chang Li[1, 2], and Fei Pang[1, 2] *

[1]*Beijing Key Laboratory of Optoelectronic Functional Materials & Micro-nano Devices, School of Physics, Renmin University of China, Beijing 100872, China*

[2]*Key Laboratory of Quantum State Construction and Manipulation (Ministry of Education), Renmin University of China, Beijing 100872, China*

*Email: feipang@ruc.edu.cn

**Abstract:**

Two-dimensional (2D) SnSe is an emerging 2D material exhibiting intriguing properties such as ferroelectricity and nonlinear optical response. Here, high-quality single-crystalline SnSe nanosheets were synthesized via NaCl-assisted chemical vapor deposition (CVD) method. The addition of NaCl was found to significantly increase the surface coverage of the nanosheets with less influence on their lateral size. The crystalline structure and composition of as-grown nanosheets were charactered by XRD, Raman spectroscopy, and XPS. Ferroelectric domains in SnSe nanosheets are directly visualized by piezoresponse force microscopy (PFM). The magnetic hysteresis loops of SnSe nanosheets are achieved at 2 K, which indicated their weak ferromagnetism. A spin glass-like behavior was observed below 115K, which is attributed to the presence of $SnSe_2$ impurity. This work further establishes a controllable synthesis route for SnSe nanosheets, thereby paving the way for subsequent investigation of their ferroelectric properties.

**Keyword:** Chemical Vapor Deposition; SnSe nanosheets; NaCl-Assisted; 2D ferroelectric materials, spin glass-like behavior

## 1. Introduction

Two-dimensional (2D) materials have attracted tremendous research interest due to their unique physical and chemical properties. Among these materials, SnSe is a member of group IV monochalcogenides due to its exceptional thermoelectric performance, anisotropic electronic structure, and phase-dependent properties.[1] In particular, few-layer SnSe exhibits ferroelectric behavior, making it a potential material for next-generation low-dimensional memory devices[2]. Furthermore, SnSe is predicted as a high mobility 2D material and possesses a layer-dependent bandgap ranging from ≈1.6 eV (monolayer) to ≈0.9 eV (bulk) with strong optical absorbance, which endows it with excellent electronic and optoelectronic performances.[3][4][5] Moreover, robust purely in-plane ferroelectricity has been observed in 2D SnSe as well as its SnS and SnTe counterparts even at monolayer thickness.[6] [7][8] More recently, studies have also suggested the possibility of inducing magnetic properties in SnSe through defect engineering, strain, or interface interactions[9]. The local phase segregation of $SnSe_2$ microdomains leads to table room-temperature ferrimagnetism in two-dimensional p-doped SnSe, [10] opening avenues for 2D multiferroic systems. The multiferroic feature makes 2D SnSe a promising candidate for nanoscale ferroelectric devices. Furthermore, spin glasses are reported in disordered systems such as dilute magnetic alloys. [11] Spin glass states are also discovered in crystalline $Cr_2Se_3$ nanosheets. [12]So to obtain high-quality and ultrathin SnSe crystals provide possible platforms for the complex magnetism physics of 2D materials.

Chemical vapor deposition (CVD) is a well-established technique for synthesizing high-quality, large-area, and thickness-controlled 2D SnSe [13][14]. Here, we design NaCl-assisted CVD approach to fabricate 2D SnSe nanosheets on mica. The introduction of NaCl promotes the vaporization of SnSe precursor during growth, which significantly increases the nucleation density of the SnSe nanosheets. The structural, ferroelectric, and magnetic properties of as-grown SnSe nanosheets were systematically investigated. Consequently, the overall surface coverage rises with increasing NaCl concentration, which has a comparatively minor effect on their lateral size of SnSe nanosheets. The room-temperature robust ferroelectricity is found by piezoresponse force microscopy (PFM) measurements. Furthermore, Magnetic measurements reveal SnSe nanosheets exhibit magnetic hysteresis loop and spin glass-like behavior.

## 2. Experimental

### 2.1 Synthesis of SnSe nanosheets

SnSe nanosheets were synthesized on fluorophlogopite mica substrates [$KMg_3(AlSi_3O_{10})F_2$] via a chemical vapor deposition (CVD) process at atmospheric pressure, as illustrated in Fig. 1a. Briefly, a mixed powder of SnSe (~60 mg) and NaCl (~3 mg) was used as the precursor and placed at the center of the high-temperature zone in a two-zone tube furnace. A freshly cleaved mica substrate was positioned approximately 9 cm downstream from the precursor. Prior to heating, the system was purged with ultrahigh-purity argon. During growth, the high-temperature zone was heated to 630 °C and the low-temperature zone to 400 °C within 15 minutes, followed by 15 minute growth time at these temperatures under a constant argon flow of 100 sccm. After growth, the furnace was cooled naturally to room temperature.

### 2.2 Characterizations

SnSe nanosheets were transferred from a mica substrate onto a $Au/SiO_2/Si$ substrate via PMMA-assisted wet transfer process. The as-grown nanosheets were characterized by optical microscopy (OM; 6XB-PC), atomic force microscopy (AFM, Bruker Dimension ICON), scanning electron microscopy (SEM, FEI NOVA NANOSEM 450) equipped with an energy-dispersive spectrometer (EDS). Their surface chemical states were analyzed by X-ray photoelectron spectroscopy (XPS, PHI-5702), and the crystalline structure was characterized by X-ray diffraction (XRD, Bruker D8 Advance, Cu-Kα radiation) and Raman spectroscopy (WITec Alpha 300R, 532 nm excitation). Magnetic properties were measured using a magnetic property measurement system (MPMS, Quantum Design), and ferroelectric property was evaluated by piezoelectric force microscopy (PFM, Park NX10).

## 3.Results and discussion

We introduce a NaCl-assisted strategy to promote growth of SnSe nanosheets. In addition, the mica substrate is also crucial for the growth of SnSe nanosheets, due to the absence of dangling bonds and relatively high thermal stability.[17][18] [19] Given the relatively high melting point of SnSe (861 °C), the addition of NaCl effectively reduced the synthesis temperature and facilitated the growth of SnSe nanosheets with well controlled thickness and lateral size. We systematically examined the role of NaCl in modulating the nucleation of SnSe nanosheets. By varying the amount of NaCl while keeping the SnSe powder mass (60 mg) and all other growth parameters constant, a clear correlation was observed between the NaCl concentration and the resulting morphology of the nanosheets.

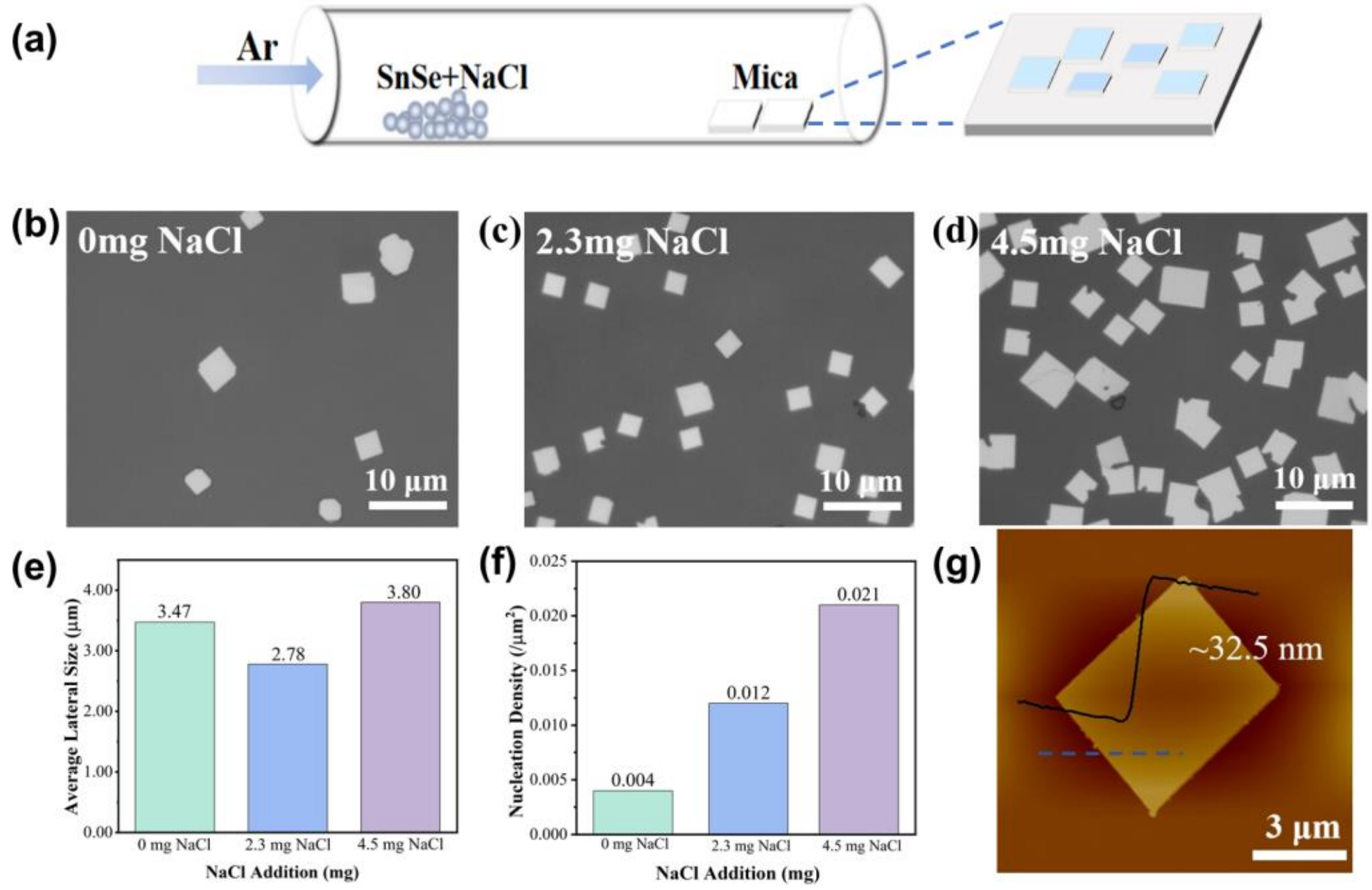

**Figure.1. Controllable synthesis of SnSe nanosheets.** (a) Schematic of the NaCl-assisted CVD setup for synthesizing SnSe nanosheets on mica. (b-d) OM images of SnSe nanosheets grown on mica with different NaCl amounts. (e-f) Evolution of the average lateral size and nucleation density of SnSe nanosheets as a function of NaCl content. (g) Typical AFM image of the as-grown SnSe nanosheet on mica.

OM images of 2D SnSe nanosheets grown with increasing amount of NaCl are shown in Fig. 1b-d, with the low magnification overview provided in Fig.S1. We performed statistical analysis on the as-grown nanosheets to evaluate the changes in lateral size and nucleation density. In the absence of NaCl, the nanosheets exhibited a low nucleation density. When 2.3 mg of NaCl was added, the lateral size of the nanosheets remained largely unchanged, while the nucleation density increased. Upon further increasing the NaCl amount to 5.4 mg, both the nucleation density and the average lateral size increased. Unlike typical NaCl-assisted growth reported in other systems, the lateral size of nanosheets here increased significantly with NaCl addition. The coverage of SnSe nanosheets was increased with amount of NaCl.

The addition of NaCl powder serves as a fluxing agent, [15][16] which lowers the melting point of the SnSe precursor and enhances its evaporation rate. Consequently, the increased vapor flux of SnSe promotes a higher nucleation density. Additionally, NaCl interacts with the SnSe precursor to form reactive intermediates that are uniformly distributed on the substrate surface. These intermediates can serve as additional nucleation sites, thereby increasing the nucleation density and improving the overall quality of the SnSe nanosheets. It should be emphasized it doesn't significant enhance the lateral growth of SnSe nanosheets with increasing NaCl concentration, while the nucleation density showed a positive correlation with the amount of NaCl (Fig. 1b-d). The morphology of SnSe nanosheets was performed using AFM. The typical AFM image is shown in Fig.1g. The thickness of the SnSe nanosheet is 32.5 nm whose surface is very smooth. The NaCl addition increased from 0 mg to 4.5 mg, the thickness of the SnSe nanosheets varied between 23 and 33 nm as shown in Fig.S2. The thickness of the nanosheets shows no clear correlation with the quantity of NaCl concentration.

As shown in Fig.2a, SnSe possesses a layered structure puckered by bimolecular $(Sn^{2+}Se^{2-})_2$ layers stacked along the *c* axis [20][21] This structure is closely related with its unique properties such as low thermal conductivity and high thermoelectric performance [6][22]. To perform XRD measurements, SnSe nanosheets were transferred onto $SiO_2$/Si substrates to avoid background interference from mica. The XRD pattern in Fig. 2b displays three main diffraction peaks, corresponding to the (200), (400), and (800) planes of SnSe, which can be indexed to the orthorhombic SnSe structure (JCPDS No. 48-1224). The predominant peaks reveal the high degree of orientation of the nanosheets and their ordered layer-by-layer stacking along the *c*-axis.

XPS was further carried out to investigate the chemical state and stoichiometry of Sn and Se. The Sn 3d and Se 3d core-level spectra are shown in Fig. 2c and 2d, respectively, and were fitted by sets of doublet peaks. Three doublet peaks (representing Sn $3d_{5/2}$ and $3d_{3/2}$) are required to fit the Sn 3d spectrum (Fig. 2c). The peaks from higher to lower binding energies can be attributed to $Sn^{4+}$ and $Sn^{2+}$,[25] respectively. The dominant doublet originates from $Sn^{2+}$, consistent with the expected state in SnSe. The $Sn^{4+}$ were likely stems from trace amounts of $SnO_2$ or $SnSe_2$ impurities present in the precursor or formed during growth. In Fig. 2d, the Se3d spectrum can be well-fitted with a set of doublet peak representing Se $3d_{5/2}$ and $3d_{3/2}$, confirming the $Se^{2-}$ in the nanosheets.

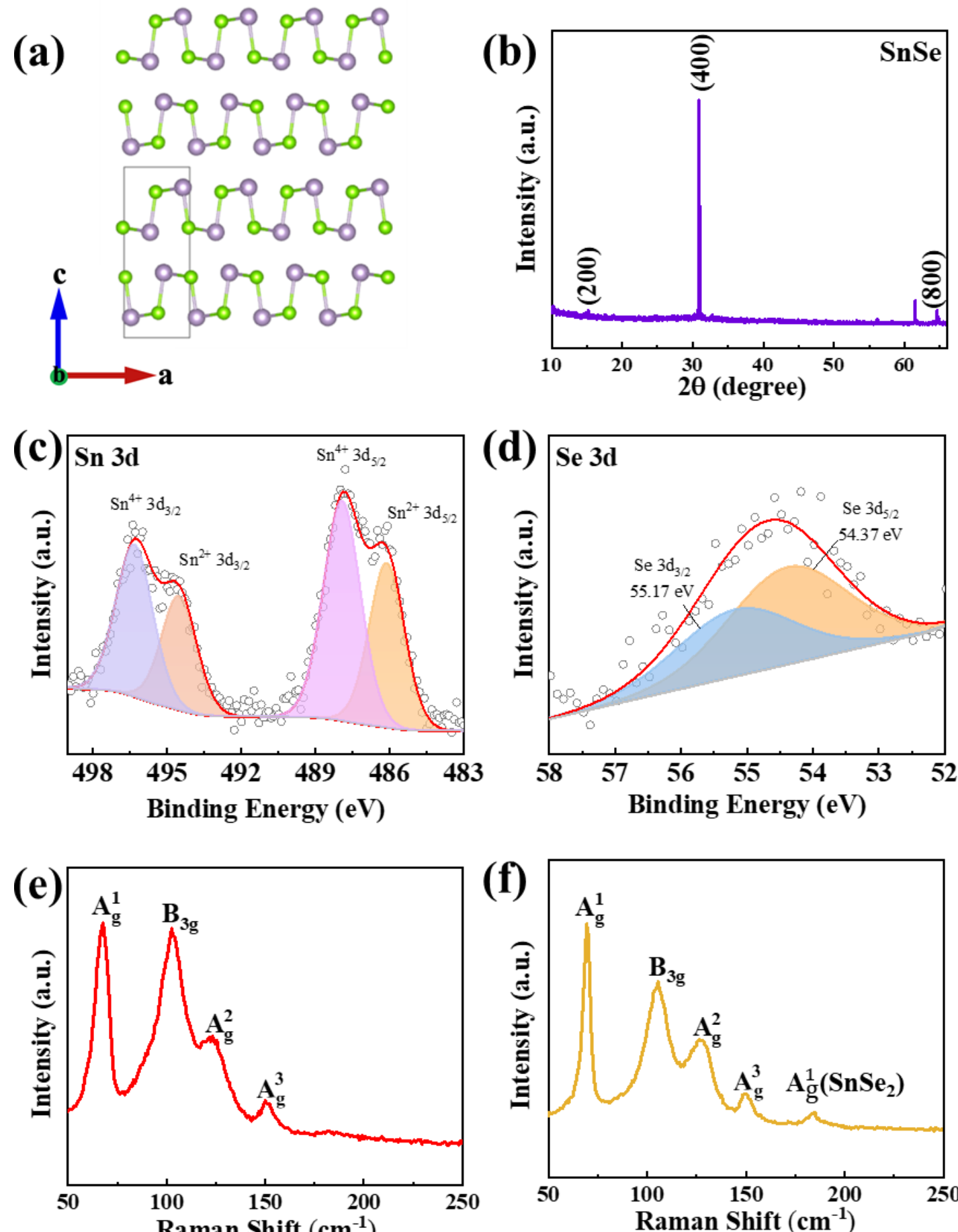

**Figure.2**. **Structural characterization of SnSe nanosheets**. (a) Side view of SnSe crystal, where green and violet balls represent the Se and Sn atoms, respectively. (b) XRD pattern of transferred SnSe nanosheets transferred onto $SiO_2$/Si substrate. (c)-(d) XPS spectra of Sn 3d (c) and Se 3d (d) orbits. (e) (f) Raman spectroscopy obtained from SnSe nanosheets without (e) and with $SnSe_2$ impurities(f).

Raman spectroscopy was used to evaluate the phase purity and crystal quality of SnSe nanosheets. As shown in Fig. 2e, the spectrum exhibits four well-resolved peaks, which correspond to the $A^1_g$, $B_{3g}$, $A_{2g}$, and $A^3_g$ vibrational modes of orthorhombic SnSe. [23][24]. No peak is observed near 476 $cm^{-1}$, which would correspond to the $E_g$ mode of $SnO_2$ [26], so $Sn^{4+}$ impurity observed in XPS is not from the $SnO_2$. However, a weak peak at ≈185 $cm^{-1}$ ($A_{1g}$) mode of $SnSe_2$ [27] was observed in some SnSe nanosheets as shown in Fig.2f. This suggests there are minor $SnSe_2$ impurities in SnSe nanosheets.[10]

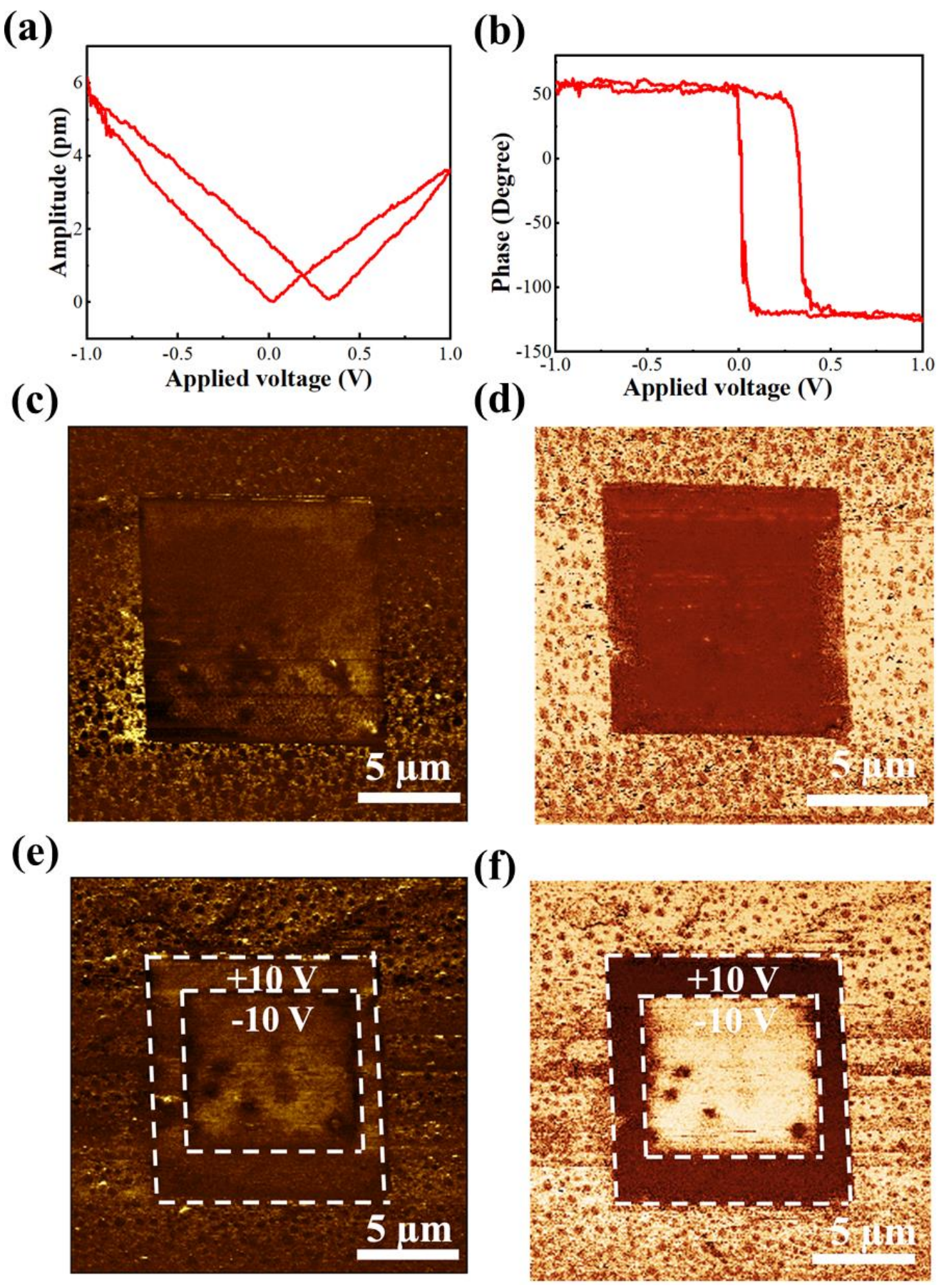

**Figure.3 Ferroelectric characterization of SnSe nanosheets.** (a), (b) PFM amplitude and phase hysteresis loops; (c), (d) amplitude and phase images of SnSe after poling with a +10 V tip bias, (e), (f) amplitude and phase images of SnSe after sequential poling with ±10 V biases

PFM is a powerful technology to investigate ferroelectricity. Here, we employed PFM to characterize SnSe nanosheets at room temperature. To eliminate the electrostatic effect, the SnSe nanosheets were transferred to the Au coated $SiO_2$/Si from mica substrates to avoid the local charge accumulation [28]. Fig. 3a-3b reveals the PFM amplitude and phase curves captured from SnSe nanosheet under an alternating bias of ±1 V. The characteristic butterfly loops of amplitude signals (Fig.3a) and the nearly 180° switching of phases (Fig.3b) clearly indicate the robust ferroelectric polarization in SnSe nanosheets. Subsequently, a square region was poled with a +10 V DC bias, and the corresponding PFM amplitude and phase images are presented in Fig. 3c and 3d. A second adjacent square was then written with a −10 V bias, producing the nested box patterns shown in the amplitude and phase images of Fig. 3e and 3f. Notably, the PFM phase image reveals bright and dark regions with a nearly 180° phase contrast, corresponding to the upwards and downwards polarization states, respectively. These results indicate the intrinsic ferroelectric nature of the SnSe nanosheets and demonstrate their reversible polarization switching.

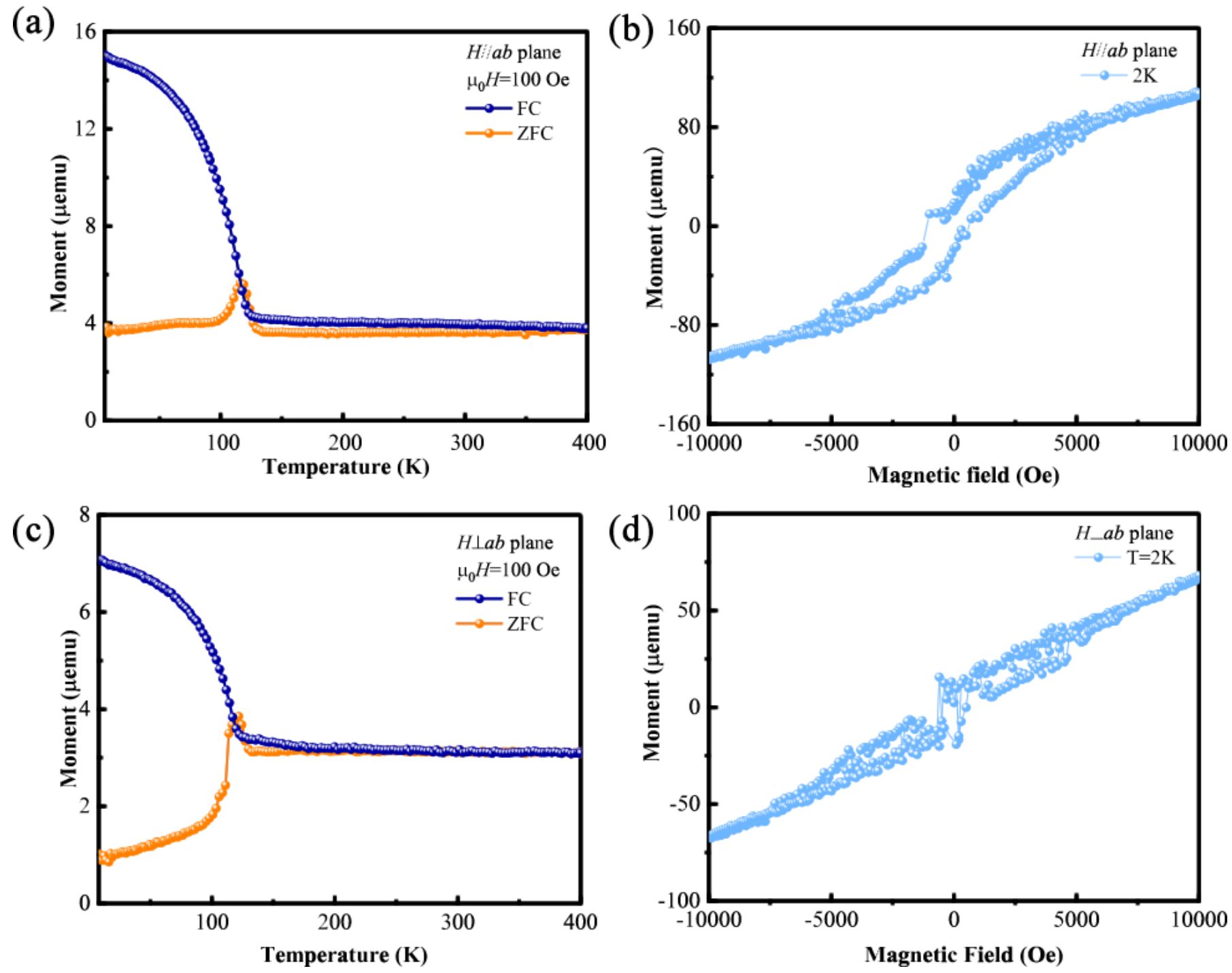

**Fig.4 Magnetic characterization of SnSe nanosheets.** (a,c) temperature-dependent magnetization (*M*-*T*) curves of SnSe nanosheets measured with a 100 Oe magnetic field applied parallel and perpendicular to *ab* plane. (b,d) magnetic hysteresis loops of SnSe nanosheets at 2 K with the field swept from −1 T to 1 T.

The ferromagnetism of 2D SnSe was firstly reported in p-doped SnSe nanosheets [10], which possibly originated from interfacial charge transfer induced by local phase segregation of $SnSe_2$ microdomains. The magnetic properties of SnSe nanosheets grown on mica were investigated by MPMS. The temperature-dependent magnetic moments measurements (*M-T*) were also performed to probe their magnetic ordering. Fig. 4a and 4c show *M-T* curves with field-cooled (FC) and zero-field-cooled (ZFC) collected with the magnetic field of 100 Oe for $H//ab$ plane and $H \perp ab$ plane, respectively. Field-cooled magnetization ($M_{FC}$) shows a paramagnetic to ferromagnetic transition at $T_c$ = 115 K and increases continuously with decreasing temperature below $T_c$. The magnetic hysteresis loops of SnSe nanosheets are achieved at 2 K under the parallel and vertical magnetic fields (Fig. 4b and 4d), which were confirm the long range ferrimagnetic order. The relatively small remanence suggested the weak ferromagnetic behavior [10]. The presence of $SnSe_2$ impurities in our samples is supported by both XPS and Raman spectra. The exchange interactions at the interfaces between the SnSe matrix and $SnSe_2$ impurities contribute to the emergence of ferromagnetic properties of SnSe Nanosheets.

Complete magnetization saturation is not observed even under applied magnetic fields of up to 1 T, is similar to other frustrated transition-metal compounds[29]. Different from $M_{FC}$, the ZFC magnetization ($M_{ZFC}$) increases with temperature below $T_c$, which is characteristic of a spin glass-like phase. This magnetic behavior possibly is attributed to the disorder induced by the phase-segregated $SnSe_2$ domains in the SnSe nanosheets. Consequently, $M_{ZFC}$ drops sharply with decreasing temperature.

## 4. Conclusions

High-crystallinity SnSe nanosheets were grown via NaCl-assisted atmospheric-pressure CVD method. The addition of NaCl promoted the vaporization of the SnSe precursor, improving the surface coverage of the 2D SnSe crystals. It was found that NaCl primarily increased the nucleation density of the SnSe nanosheets, while exerting less influence on their lateral size. Room-temperature ferroelectricity in the nanosheets was confirmed by PFM, with nested box patterns being successfully written under opposite bias voltages. Moreover, the nanosheets exhibited weak ferromagnetism along with a spin glass-like state, which is attributed to the presence of $SnSe_2$ impurities. The NaCl-assisted CVD strategy developed in this work offers a way for synthesizing 2D SnSe, thereby establishing a promising material platform for exploring potential applications in 2D multiferroics.

**Acknowledgements**

This work was supported by the National Natural Science Foundation of China (NSFC) (No. 12374200 and 92477205).